\documentclass[preprint,showpacs,amsmath,amssymb]{revtex4}
\usepackage{graphicx}
\usepackage{amsmath}
\usepackage{bm}
\usepackage{relsize}
\RequirePackage{xspace}
\newcommand{\babar}{\mbox{\slshape B\kern-0.1em{\smaller A}\kern-0.1em %
B\kern-0.1em{\smaller A\kern-0.2em R}}\xspace}
\def\bea{\begin{eqnarray}}
\def\eea{\end{eqnarray}}
\def\nn{\nonumber}

\begin{document}

\title{\mbox{}\\[10pt]
 Lepton Flavor Violation as a Probe of Quark-Lepton Unification}

\author{ Kingman Cheung }
\address{ Department of Physics, Tsing Hua Univeristy, Hsinchu,
Taiwan}

\author{Sin Kyu Kang}
\affiliation{ School of Physics, Seoul National University, Seoul
151-741, Korea}

\author{C. S. Kim and Jake Lee}
\affiliation{ Department of Physics, Yonsei University, Seoul
120-749, Korea}



\date{\today}
\begin{abstract}
\noindent The recent measurements of the solar neutrino mixing angle
$\theta_{sol}$ and the Cabibbo mixing angle $\theta_C$ reveal a
surprising relation, $ \theta_{sol}+\theta_C\simeq\frac{\pi}{4} $.
Interpreting this empirical relation as a support of the
quark-lepton unification, we find that the PMNS mixing matrix
can be decomposed into
a CKM-like matrix and maximal mixing matrices, which
can give  profound
implications on the quark-lepton unification.
We explore a possibility to probe the implication of quark-lepton
unification by considering the relative sizes of branching ratios for the lepton
flavor violating radiative decay processes, $l_i\rightarrow
l_j\gamma$, in the context of the supersymmetric standard model
with heavy right-handed Majorana neutrinos.

\end{abstract}

%

\maketitle

Neutrino studies will enter a new era when the MINOS experiment
starts firing a neutrino beam toward the Soudan mine in March
2005. Until now, while the atmospheric neutrino deficit still
points toward a maximal mixing between the tau and muon neutrinos,
however the solar neutrino problem favors a not-so-maximal mixing
between the electron and muon neutrinos.  Surprisingly, it has
recently been noted that the solar neutrino mixing angle
$\theta_{sol}$ required for a solution of the solar neutrino
problem and the Cabibbo angle $\theta_C$ reveal a striking
relation \cite{raidal} \bea \theta_{sol}+\theta_C \simeq
\frac{\pi}{4}, \eea which is satisfied by the experimental results
within a few percent accuracy
$\theta_{sol}+\theta_{C}=45.4^{\circ}\pm 1.7^{\circ}$
\cite{SK2002,SNO,fits}. This quark-lepton complementarity (QLC)
relation (1) has been simply interpreted  as an evidence for
certain quark-lepton symmetry or quark-lepton unification as shown
in Refs. \cite{raidal,smirnov,mohap,kkl}.

To effectively describe the deviation from maximal mixing of solar
neutrino  as well as a small mixing element $U_{e3}$ and possible
deviation from maximal mixing of atmospheric neutrino, three
possible combinations of maximal mixing and a certain mixing
matrix $U(\lambda)$ parameterized in terms of a small parameter
$\lambda\sim \sin\theta_C$ have been proposed as parametrization
of $U_{\rm PMNS}$ \cite{giunti, rodej, frampton, ramond}:
 \bea
(\mbox{a})& & U^{\dagger}(\lambda)U_{\rm bimax}, \nn \\
(\mbox{b})& & U_{\rm bimax}U^{\dagger}(\lambda), \label{framp} \\
(\mbox{c}) & & U^m_{23}U^{\dagger}(\lambda)U^m_{12}. \nn
\eea
Here $U_{\rm bimax}$ corresponds to the bi-maximal lepton mixing matrix
\cite{bimax}, and $U_{23}^m,U_{12}^m$ denote the rotation matrices
with (2,3) and (1,2) maximal mixing, respectively. Even though the
present data is not sufficient to determine which combination can
give correct flavor structure in lepton sectors, it is very
important to single them out because the QLC relation is strongly
correlated to each combination differently. As
extensively studied in \cite{smirnov}, the QLC relation can be
derived from the parametrization given above but up to some
corrections. These corrections to the QLC relation can be
compensated with renormalization effects \cite{kkl}.

In this Letter, we will show that among possible forms of
$U(\lambda)$ which are consistent with the neutrino experimental results,
the ``CKM-like" form of $U(\lambda)$ has profound implication on
the quark-lepton unification. Motivated by this observation, we
will study the implication of the parametrization composed of
bi-maximal mixing and $U_{\rm CKM}$ reflecting quark-lepton
unification by considering the lepton flavor violating (LFV)
decays particularly in the context of supersymmetric standard
model (SSM). We also examine a possibility to differentiate the
above combinations by considering the relative size of branching
ratios of the radiative LFV decays, $Br(l_i\rightarrow l_j\gamma)
~ (i,j=e,\mu,\tau)$. While the LFV processes have tiny rates in
the minimal extensions of the standard model (SM) with heavy
right-handed Majorana neutrinos, the supersymmetric extensions of
the SM can lead to sizable effects on the LFV processes due to new
sources of lepton flavor violation. As is well known, the LFV
decays in SSM can be caused by the misalignment of lepton and
slepton mass matrices \cite{masiero} and the branching ratios of
the LFV decays depend on the specific structure of the neutrino
Dirac Yukawa matrix $Y_{\nu}$. Therefore, we expect that a
specific structure of $Y_{\nu}$ reflecting quark-lepton
unification can lead to distinctive predictions for the branching
ratios of the LFV decays. However, the branching ratios of the LFV
decays in SSM strongly depend on several parameters which make it
difficult to probe the structure of $Y_{\nu}$. Instead of
considering the branching ratios of each LFV process, we can rely
on the relative size of $Br(l_i \rightarrow l_j \gamma)$ among the
three different flavors, because the relative size is almost free
from arbitrary supersymmetric parameters. These ratios of $Br(l_i
\rightarrow l_j\gamma)$ can be useful to probe the structure of
$Y_{\nu}$ with the help of the parametrization of $U_{\rm PMNS}$
given in Eq.~(\ref{framp}). In particular, we expect that a
hierarchical structure of $Y_{\nu}$ predicted by quark-lepton
unification may be responsible for the hierarchy of
$Br(l_i\rightarrow l_j\gamma)$ if they are observed in the future
experiments. In such a way, the quark-lepton unification could be
tested from the determination of the relative size of the
branching ratios in future experiments.

Let us begin by considering how the parametrization given by the
forms of Eq.~(\ref{framp}) can be realized in the framework of the
quark-lepton unification. For our purpose, it is useful to work in
a basis where the quark and lepton Yukawa matrices are related to
each other by a certain symmetry. In general, the quark Yukawa
matrices $Y_u,Y_d$ are given by $
Y_u=U_uY_u^{diag}V_u^{\dagger},~~Y_d=U_dY_d^{diag}V_d^{\dagger},$
from which the observable CKM quark mixing matrix is described as
$ U_{\rm CKM}=U^{\dagger}_uU_d$. For the lepton sector, we
consider the following leptonic superpotential, which implements
the seesaw mechanism:
  \bea W_{\rm
lepton}=Y_l\widehat{L}\widehat{l}_L^c\widehat{H}_d
+Y_{\nu}\widehat{L}\widehat{N}^c_L\widehat{H}_u
-\frac{1}{2}\widehat{N}^{cT}_LM_R\widehat{N}^c_L,
\label{lagrang}\eea where the family indices have been suppressed
and $\widehat{L}_j$, $j=e,\mu, \tau\equiv 1,2,3$, represent the
chiral super-multiplets of the $SU(2)_L$ doublet lepton fields,
$\widehat{N}_{jL}^c, \widehat{l}^c_{jL}$ are the super-multiplet
of the $SU(2)_L$ singlet neutrino and charged lepton field,
respectively. In the superpotential $W_{\rm lepton}$, $M_R$ is the
heavy Majorana neutrino mass matrix. $Y_l$ and $Y_{\nu}$ are the
$3\times 3$ charged lepton and neutrino Dirac Yukawa matrices,
respectively and can be parameterized as \bea Y_l =
U_lY_l^{diag}V^{\dagger}_l, ~~ Y_{\nu} =
U_0Y^{diag}_{\nu}V^{\dagger}_0.\label{leptonY}\eea
We note that in
the framework of the minimal unification and the symmetric basis
where the quarks and leptons are interrelated, $M_R$ is generally
not diagonal. The light neutrino mass matrix can be generated
through the seesaw mechanism after the breaking of the electroweak
symmetry as
\bea M_{\nu}
= \left( U_{0}M_{\rm Dirac}^{diag}V^{\dagger}_0 \right) M_R^{-1}
\left(V^{\ast}_0 M_{\rm Dirac}^{diag} U^T_0 \right),\label{lep1}
\eea where $M_{\rm Dirac}=Y_{\nu}v_u/\sqrt{2}$ with
$v_u=v\sin\beta$. We can then rewrite $M_{\nu}$ as follows \bea
M_{\nu} = U_0V_M M_{\nu}^{diag}V_M^T U_0^T, \label{lep2}\eea where
$V_M$ represents the diagonalizing matrix of
\bea M_{\rm Dirac}^{diag}V^{\dagger}_0 M_R^{-1}
     V^{\ast}_0 M_{\rm Dirac}^{diag}. \nn \eea
Then, the observable PMNS mixing matrix can be written as
\bea
U_{\rm PMNS} = U^{\dagger}_l U_{\nu} =
U^{\dagger}_lU_0V_M.\label{lep3} \eea

 Now, let us consider how
$U_{\rm PMNS}$ given by Eq.~(\ref{lep3}) can be related with
$U_{\rm CKM}$ in the context of quark-lepton unification. The
quark-lepton unification based on the minimal $SU(5)$ leads to the
following simple relations,
\bea Y_e=Y^T_d,~~~Y_u=Y^T_u. \eea
Then, we deduce that
  $U_l=V_d^{\ast}$ from which
  \bea
  U_{\rm PMNS}=V^T_d U_0 V_M.\eea
As one can easily see, the contribution of $U_{\rm CKM}$ may
appear in $U_{\rm PMNS}$ if we further assume that the Yukawa
matrix of the up-type quark sector is related with that of
Dirac-type neutrinos such as $Y_{\nu}=Y_u$ which can be realized
in some larger unified gauge group such as $SO(10)$. Then, the
lepton flavor mixing matrix can be written as \bea U_{\rm PMNS} =
V^T_d U_d U_{\rm CKM}^{\dagger}V_M. \label{pmnsG}\eea In addition,
requiring symmetric form of the down-type quark Yukawa matrix, we
obtain \bea U_{\rm PMNS}=U^{\dagger}_{\rm CKM}V_M, \label{sym}\eea
 where the mixing
matrix $V_M$ should have two almost maximal mixings so as to
account for the solar and atmospheric neutrino oscillations. This
expression for $U_{\rm PMNS}$ corresponds to the parametrization
given by Eq.~(\ref{framp}-a). On the other hand, in order to
achieve the parametrization given by Eq.~(\ref{framp}-b), one
should take $V_M$ to be identity matrix and the product of two
matrix $V^T_dU_d$ in Eq.~(\ref{pmnsG}) should give bi-maximal
mixing pattern. Since the left-handed rotation matrix $U_d$ for
down type quark can be almost diagonal to leading order,
$V^T_d$ should have almost bi-maximal mixing form, which can be
achieved in the so-called lopsided form of Yukawa matrix. The case
given by Eq.~(\ref{framp}-c) can also be achieved by taking
$V^T_d\simeq U_{23}^m$ and $V_M\simeq U_{12}^m$ in Eq.~(\ref{pmnsG}).
In such ways, $U_{\rm PMNS}$ can be connected with
$U_{\rm CKM}$ in the framework of the quark-lepton unification.

Although the minimal quark-lepton unification can lead to an
elegant relation between $U_{\rm PMNS}$  and $U_{\rm CKM}$ as shown
above, it indicates undesirable mass relations between quarks and
leptons at the GUT scale such as $m_d^{diag}=m_l^{diag}$. Thus, we
need to modify the simple relations between quark and lepton
Yukawa matrices so as to achieve desirable mass relations. {}From
the well known empirical relation
\bea |V_{us}| \simeq
\sqrt{\frac{m_d}{m_s}}\simeq 3\sqrt{\frac{m_e}{m_{\mu}}},
\label{emp}
\eea
it has been shown that the $U(\lambda)$ in Eq.
(\ref{framp}) should have the CKM-like form but with the
replacement of $\lambda$ with $\lambda/3$ as shown in
Refs.~\cite{pakvasa,kkl}, which can be obtained by introducing the
Higgs sector transforming under the representation {\bf 45} of
$SU(5)$ or {\bf 126} of $SO(10)$ \cite{higgs}.

Now, let us consider how the relative ratio of $Br(l_i\rightarrow l_j
\gamma)$ can be connected with the
structure of the neutrino Dirac Yukawa matrix, which is constructed
from the grand unification scenario above. It is well known that
the RG running induces off-diagonal terms in the slepton mass
matrix even for the case of universal slepton masses at GUT scale
\cite{casas}:
\bea m^2_{\tilde{l}_{ij}}\simeq
-\frac{1}{8\pi^2}(3m_0^2+A_0^2)(Y^{\prime}_{\nu}Y^{\prime\dagger}_{\nu})_{ij}
\log\frac{M_G}{M_X}, \label{slept}
\eea
where $m_0, A_0$ are
universal soft scalar mass and soft trilinear $A$ parameter, and
$Y^{\prime}_{\nu}$ is defined in the basis where the charged
lepton Yukawa matrix and the heavy Majorana mass matrix are real
and diagonal. Here $M_G$ and $M_X$ denote the GUT scale and the
characteristic scale of the right-handed neutrinos at which
off-diagonal contributions are decoupled \cite{casas},
respectively. Thus, one can expect that some specific form of
$Y^{\prime}_{\nu}$ is crucial to estimate the sizes of LFV
processes which are caused by non-diagonal slepton mass matrix.
First of all, let us consider the parametrization (a). It follows
from Eqs.~(\ref{framp}-a,\ref{sym}) that \bea Y^{\prime}_{\nu}
\equiv U^{\dagger}_{\rm CKM}Y^D_{\nu}V_{0}^TV_R,\label{yuk}\eea
where $Y^D_{\nu}$ is diagonal neutrino Dirac Yukawa matrix and
$V_R$ is the rotation matrix of the heavy Majorana neutrino mass
matrix $M_R$ in Eq.~(\ref{lagrang}). {}From Eq.~(\ref{yuk}), the
term $(Y_{\nu}^{\prime}Y^{\prime\dagger}_{\nu})$ becomes
\bea
Y_{\nu}^{\prime}Y^{\prime \dagger}_{\nu}=U^{\dagger}_{\rm
CKM}(Y^D_{\nu})^2U_{\rm CKM}. \label{yuk2}
\eea

 The induced off-diagonal terms in the
slepton mass matrix can be a source of the lepton flavor violation
in SSM and they can yield sizable contributions to LFV decays,
$l_i\rightarrow l_j \gamma$. The contribution to the branching
ratios of the LFV decays due to the slepton mass term is roughly
given by \bea Br(l_i\rightarrow l_j \gamma)\simeq
\frac{\alpha^3}{G^2_F}\tan^2\beta\left|\frac{m^2_{\tilde{l}_{ij}}}{m^4_S}\right|^2,
\label{br} \eea where $m_S$ is a supersymmetric mass scale. Let us
define $Y^D_{\nu}$ in a hierarchical form expressed in terms of
the power of $\lambda$ : \bea Y^D_{\nu}\equiv
Y_3\left(\begin{array}{ccc} \lambda^{n_1} & &
\\ & \lambda^{n_2} & \\& & 1 \end{array} \right). \label{hier}\eea
For the quark-lepton unification, $Y_3=m_t/v_u$ and the powers of
$\lambda$ are given by $n_1=8, n_2=4$ so as to be the same
hierarchy of up-type quark sector at high energy scale.

The term $(Y_{\nu}^{\prime}Y^{\prime \dagger}_{\nu})$ is roughly
given to leading order as \bea
&&Y_{\nu}^{\prime}Y^{\prime\dagger}_{\nu} \sim
\left(\frac{m_t}{v_u}\right)^2\times  \label{term2}\\
&& \left(\begin{array}{ccc}
 \lambda^{2n_1}+\lambda^{2n_2+2}+\lambda^6 & \lambda^{2n_1+1}-\lambda^{2n_2+1}
 -\lambda^5 & \lambda^3 \\
 \lambda^{2n_1+1}-\lambda^{2n_2+1}-\lambda^5 & \lambda^{2n_1+2}+\lambda^{2n_2}
 +\lambda^4 & -\lambda^2 \\
 \lambda^{3} & -\lambda^{2}
& 1
\end{array}\right). \nn \eea
Inserting this into Eq.~(\ref{br}), we can estimate how large
$Br(l_i\rightarrow l_j\gamma)$ could be by fixing the parameters
$m_S$ and $M_X$.
Instead of considering the values of
$Br(l_i\rightarrow l_j\gamma)$, we focus on the ratio of
$Br(l_i\rightarrow l_j\gamma)$. The ratio of $Br(l_i\rightarrow
l_j\gamma)$ only depends on $Y^{\prime}_{\nu}Y^{\prime
\dagger}_{\nu}$, and thus from Eqs.~(\ref{slept},\ref{br},\ref{term2}),
we can simply obtain the ratio:
\bea Br(\mu\rightarrow e \gamma):Br(\tau\rightarrow e
\gamma):Br(\tau\rightarrow \mu\gamma) \nn \\ \simeq ~~~
(-\lambda^{2n_1-1}+\lambda^{2n_2-1}+\lambda^{3})^2:\lambda^{2}:1 .
\label{case1}
\eea
When $Y^D_{\nu}$ is the same as $Y^D_u$ due to the quark-lepton
unification, we predict that the ratio given in Eq.~(\ref{case1})
should be $(\lambda^6:\lambda^2:1)$.
But, we note that this ratio may not necessarily indicate
quark-lepton unification just considered because we can obtain the
same ratio in the limit of large values of $n_1, n_2$. However, if
the ratio of $Br(l_i\rightarrow l_j\gamma)$ is measured to be
inconsistent with the prediction of the ratio given above,
it may indicate $Y^D_{\nu}\neq Y^D_u$.

For the case of realistic quark-lepton
unification satisfying Eq.~(\ref{emp}),
 the terms
$(Y_{\nu}^{\prime}Y^{\prime \dagger}_{\nu})_{i,j}$ for
$(i,j)=(2,1),(3,1),(3,2)$ are roughly given as \bea
(Y_{\nu}^{\prime}Y^{\prime \dagger}_{\nu})_{21} &\simeq&
\frac{\lambda^5}{6}+\frac{\lambda^{1+2n_1}}{3}-\frac{\lambda^{1+2n_2}}{3} \nn \\
(Y_{\nu}^{\prime}Y^{\prime \dagger}_{\nu})_{31} &\simeq&
\frac{\lambda^3}{6}-\frac{\lambda^{3+2n_1}}{2}+\frac{\lambda^{3+2n_2}}{3} \nn \\
(Y_{\nu}^{\prime}Y^{\prime \dagger}_{\nu})_{32} &\simeq&
\lambda^2-\frac{\lambda^{4+2n_1}}{6}-\lambda^{2+2n_2}.  \eea Then,
the ratio of $Br(l_i\rightarrow l_j\gamma)$ among the three different
flavors is
\bea && Br(\mu\rightarrow e \gamma):Br(\tau\rightarrow
e \gamma):Br(\tau\rightarrow \mu\gamma) \nn \\ && \simeq ~~~
(\lambda^{2n_1}-\lambda^{2n_2}+\lambda^{4})^2:\lambda^{4}:1,
 \label{case2}
 \eea
in order of magnitude estimation. For $n_1=8, n_2=4$, the ratio
becomes $\lambda^8:\lambda^4:1$. Therefore, we may
confirm the validity or breaking of the quark-lepton unification
through the measurements of the ratios of $Br(l_i\rightarrow
l_j\gamma)$. As can be seen from Eq.~(\ref{term2}), the important
elements in $U_{\rm CKM}$ which actually determine the hierarchy
among $Br(l_i\rightarrow l_j\gamma)$ are $(U_{\rm CKM})_{13}$ and
$(U_{\rm CKM})_{23}$. For more precise predictions of the relative
branching ratios, it is urgently required to determine $(U_{\rm
CKM})_{13}$ and $(U_{\rm CKM})_{23}$ experimentally with better
accuracy.
Note that the lepton mixing matrix given by Eq.~(\ref{sym}) leads
to a new QLC relation,
\bea (U_{\rm PMNS})_{e3}=[-\lambda+(U_{\rm
CKM})_{31}]/\sqrt{2}.
\eea
Therefore, we are led to further
confirm or discard quark-lepton unification through the
measurement of $(U_{\rm CKM})_{31}$ and $(U_{\rm PMNS})_{e3}$.
Note that similar to the QLC relation between $\theta_{sol}$ and $\theta_C$,
we can get another QLC relation between the mixing angle $\theta_{atm}$
and $(\theta_{23})_{\rm CKM}$,
\bea
\theta_{atm}+(\theta_{23})_{\rm CKM} \simeq \pi/4.
\eea

Similar to the parametrization (\ref{framp}-a), we can easily
estimate the relative ratios of $Br(l_i\rightarrow l_j\gamma)$ for the parameterizations
in (\ref{framp}-b) and (\ref{framp}-c). In these cases, the term
$Y^{\prime}_{\nu}Y^{\prime \dagger}_{\nu}$ becomes
\bea
Y^{\prime}_{\nu} Y^{\prime\dagger}_{\nu}=\left\{
\begin{array}{ll} U_{\rm bimax}U^{\dagger}_{\rm
CKM}(Y^D_{\nu})^2U_{\rm CKM}U^{\dagger}_{\rm bimax} & (\mbox{2-b}), \\
U_{23}^m U^{\dagger}_{\rm CKM}(Y^D_{\nu})^2 U_{\rm CKM}U^{m
\dagger}_{\rm 23} & (\mbox{2-c}). \end{array} \right.
\eea
Imposing
the hierarchy of $Y^D_{\nu}$ given by Eq.~(\ref{hier}), the relative ratios
of $Br(l_i\rightarrow l_j\gamma)$
become
\bea Br(\mu\rightarrow e \gamma):Br(\tau\rightarrow e
\gamma):Br(\tau\rightarrow \mu\gamma) \nn \\ \simeq ~~~
\lambda^4:\lambda^4:1 ~~\mbox{(2-b)}, ~~~~~ \lambda^6:\lambda^6:1
  ~~\mbox{(2-c)}.
  \label{case3}\eea
{}From the predictions (\ref{case1},\ref{case3}), one can see that
experimental determination of the relative ratios of
$Br(l_i\rightarrow l_j\gamma)$ can differentiate the
parameterizations of the quark-lepton unification
if the empirical QLC relations indeed indicate the quark-lepton unification.

We note that the RG-induced off-diagonal terms in the slepton mass
matrix is more precisely given by \cite{ellis} \bea
m^2_{\tilde{l}_{ij}}\simeq
-\frac{1}{8\pi^2}(3m_0^2+A_0^2)\left(Y^{\prime}_{\nu
ik}\log\frac{M_G}{M_{R_k}}Y^{\prime\dagger}_{\nu kj}\right).
\label{slept2} \eea In this expression, we see that the prediction
of $Br(l_i\rightarrow l_j\gamma)$ depends on the hierarchy of the
heavy Majorana neutrino mass eigenvalues $M_{R_k}$.
However, we note that the hierarchy is not
arbitrary but derived from seesaw formulae if we fix a light
neutrino mass $m_{\nu_1}$ in our framework. According to our
numerical estimation on the relative ratios of $Br(l_i\rightarrow
l_j\gamma)$ based on Eq.~(\ref{slept2}), the hierarchical patterns
given in Eqs.~(\ref{case1},\ref{case2},\ref{case3}) are almost
kept, as long as $m_{\nu_1} \geq 10^{-5}$ eV.
This is due to the mild hierarchy of $\log\frac{M_G}{M_{R_k}}$.

In summary,
interpreting the surprising empirical relation,
$\theta_{sol}+\theta_C \simeq \frac{\pi}{4}$, as a support of the
quark-lepton unification, we find that the PMNS mixing matrix
can be parameterized by a CKM-like matrix and maximal mixing matrices
in various ways.
Each parametrization may imply very
different fundamental flavor structure. We have shown that the
various parameterizations of $U_{\rm PMNS}$ with regard to
quark-lepton unification would give very different and profound
implication to the radiative leptonic decays, $l_i \rightarrow
l_j\gamma$, in the context of SSM.
Therefore, by measuring the relative size of the
radiative decay branching ratios, we will be able to pin down the
$U_{\rm PMNS}$ parametrization, assuming the quark-lepton
unification. There have been proposed experiments
\cite{experiments} to measure these radiative decays.
The proposal in this Letter can soon be tested for the
quark-lepton unification.
\\

S.K.K and J.L. are supported  by BK21 program of the Ministry of Education
in Korea. C.S.K. is supported by Grant
No.R02-2003-000-10050-0 from BRP of the KOSEF.
K.C. is supported by the NSC of Taiwan.

\end{document}